\documentclass[runningheads]{llncs}
\usepackage[T1]{fontenc}
\usepackage{hyperref}
\usepackage{graphicx}
\begin{document}
\title{Text2Playlist: Generating Personalized Playlists from Text on Deezer}
%\title{Text2Playlist: Generating Personalized Playlists from Text on Deezer}
%
\titlerunning{Text2Playlist}
%\author[1]{Mathieu Delcluze}[%
%email=research@deezer.com
%]
%\address[1]{Deezer Research, Paris, France}

\author{Mathieu Delcluze, Antoine Khoury, Clémence Vast, Valerio Arnaudo, \\ Léa Briand,  Walid Bendada, \and Thomas Bouabça}

\authorrunning{M. Delcluze, et al.}

\institute{Deezer Research, Paris, France \\
\email{research@deezer.com}
}

\maketitle              % typeset the header of the contribution
\begin{abstract}
%Although the search engine of the streaming service Deezer is essential to browse and retrieve items from its large music catalog, it is not primarly designed for extensive exploration. Text2Playlist leverages recent advances in generative AI, combined with music information retrieval and recommendation systems, to create personalized playlists from text at scale. We detail the development and the deployment of this system on Deezer.

The streaming service Deezer heavily relies on the search to help users navigate through its extensive music catalog. Nonetheless, it is primarily designed to find specific items and does not lead directly to a smooth listening experience. We present Text2Playlist, a stand-alone tool that addresses these limitations. Text2Playlist leverages generative AI, music information retrieval and recommendation systems to generate query-specific and personalized  playlists, successfully deployed at scale. %We detail the motivations, the development, and the deployment of this system on Deezer.

\keywords{Generative AI \and Music Information Retrieval \and Real-World Deployment}
\end{abstract}

\section{Introduction}

\label{section_intro}

Search engines of online content platforms are essential to explore large catalogs of items \cite{lamkhede2019challenges}. Traditionally, search systems have been optimized for \textit{narrow} intent queries, where users have a \textit{focus} mindset aimed at a \textit{navigational} goal \cite{sguerra2022navigational,hosey2019just}, i.e., looking for specific entities such as music tracks, products and books \cite{penha2023improving,bainbridge2003people,garcia2018understanding,laplante2008everyday}. In contrast, users with an \textit{exploratory} mindset, aimed at an \textit{informational} goal, use \textit{broad} intent queries (e.g., in music domain “Chill vibes on a rainy afternoon”) \cite{sguerra2022navigational,li2019search,su2018user}. In particular, the search engine of the French music streaming service Deezer \cite{deezerwebsite} helps 16 million users from 180 countries access more than 120 millions of music tracks. Despite supporting both intent queries, its design prioritizes navigation: the small tool bar does not encourage long queries and search results cannot be easily transformed into playlists. 

\begin{figure*}[ht]
  \centering
  \fbox{\includegraphics[width=0.8\linewidth]{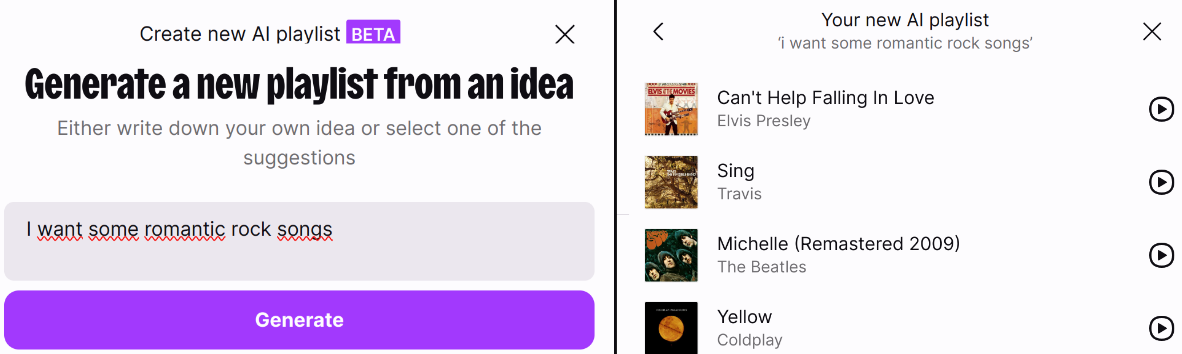}}
  \caption{Interface of the ``Text2Playlist'' tool on the website version of Deezer, creating a personalized playlist generated from an idea given by a text from a Deezer user.}
\label{text2playlistfigure}
\end{figure*}

In this paper, we present Text2Playlist, illustrated in Figure \ref{text2playlistfigure}, a personalized playlist creation tool, explicitly dedicated for broad intent queries, distinct from the search feature. It takes advantage of the recent rise of Large-Language Models (LLMs) \cite{zhao2023survey,zhou2023comprehensive,petroni2019language} and gets inspiration from Retrieval-Augmentation Generation (RAG) framework \cite{lewis2020retrieval}. Text2Playlist has been deployed on Deezer mobile and web applications for a first test phase of 5\% of premium users since July 2024 and 20\% since October 2024. %, helping the creation of more than XXX personal playlists - CONFIDENTIAL.
This paper is organized as follows. In Section \ref{context}, we further detail our motivations for such a system. In Section \ref{text2playlistsection} we detail how we use together query extraction, various metadata sources, Deezer recommendation system and LLMs to build this solution. In Section \ref{deploy}, we explain how Text2Playlist is deployed on Deezer and present analysis based on the first data gathered. We conclude and discuss areas of improvement in Section \ref{conclusion}.

\section{Where to Explore and Save Music on Deezer}
\label{context}

To explore the Deezer catalog, users can either rely on recommendation features or use the search function for more targeted music content. In this paper, we will focus on this latter way to explore and save music. Intent of the query - narrow vs broad - is identified by the search system thanks to a model combining search patterns and item clicks among historical queries \cite{sguerra2022navigational,li2019search}. On Deezer, narrow queries comprise 80\% of the total; they focus on providing the specific music entity specified in the query - it can be tracks, artists or albums. Although the remaining 20\% of broad queries are less numerous and more challenging to satisfy, they are essential to address as they foster further catalog exploration and enrichment of personal libraries, both strong signals of engagement on the platform \cite{hosey2019just}. Despite these promising results, the current search bar is intentionally primarily designed for narrow queries and is not adapted to long, complex - broad queries. Besides, it is not straightforward to transform search results into a seamless music experience. Indeed, before listening to the search output, users must manually sort content into playlists, which can be time-intensive.

\section{Text2Playlist, a Playlist Generation Tool from Text}
\label{text2playlistsection}

\begin{figure*}[ht]
  \centering
  \includegraphics[width=1\textwidth]{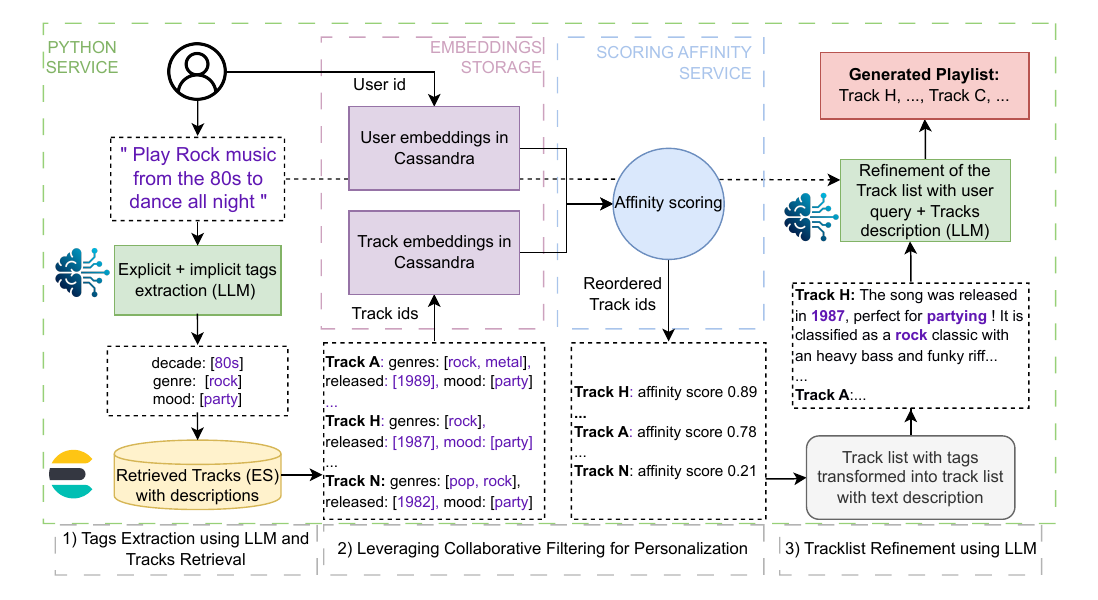}
  \caption{Overview of the Text2Playlist framework from Section \ref{text2playlistsection}, available for online requests in our production environment on Deezer.}
\label{figure-overview}
\end{figure*}

Developed in 2024, the Text2Playlist engine, illustrated in Figure \ref{text2playlistfigure}, aims to address these limitations. It encourages users to write their music needs and generates a personalized playlist, tailored to the query. Figure \ref{figure-overview} provides a summary of Text2Playlist system in production, further described in this section.

%\subsection{Retrieving a first list of tracks matching the music needs}
\subsection{Tags Extraction using LLM and Tracks Retrieval}
\label{tagsandretrieval}
Tags are often characterized as keywords to describe key information (e.g., for music items we can talk about music genre, decade, mood, artist gender, language...). They are useful to boost relevance matching, help query reformulation and item recommendation \cite{lee2017survey,li2017leveraging,mao2020item}. A query may contain information that is explicitly or implicitly expressed. For example, in the query “I want music from the 90s for work”, we can extract the explicit decade tag “90s” named by the user, but we can also infer they may prefer “Focus” mood tracks to be able to work during their listening session. Thanks to a LLM, we deduce both explicit and implicit tags from the query \cite{bai2024leveraging,issa2023comparative,naveed2023comprehensive}. Then we leverage tags already existing to describe the Deezer catalog. It comes from manual annotations from music experts, but also various internal models relying on audio content analysis and user-made playlists to expand the coverage \cite{meseguer2024experimental,gong2021ast,bontempelli2022flow,ibrahim2022exploiting,ibrahim2020should,epure2020multilingual,choi2020prediction,ibrahim2020audio,hennequin2018audio,delbouys2018music}. Tracks matching the extracted tags from the query using the LLM are retrieved and factored into JSON format.

\subsection{Leveraging Collaborative Filtering for Personalization}

\label{leveragingcf}

Most of our personalized recommender systems~\cite{briand2021semi,bendada2020carousel,bendada2023track,bontempelli2022flow,briand2024let} leverage latent models for Collaborative Filtering (CF)~\cite{bokde2015matrix,koren2015advances}. By analyzing usage data on Deezer, they learn low-dimensional \textit{embedding} vector representations of users and tracks, in a vector space where proximity reflects~preferences.
Then, they offer recommendations based on embedding similarity metrics~\cite{briand2021semi,bontempelli2022flow}. After computing the cosine similarities between the user and the tracks obtained in \ref{tagsandretrieval}, the list is reordered in descending order of similarity, first tracks being the closest to the user profile.

\subsection{Tracklist Refinement using LLM}

\label{playlistrefinement}

Inspired by two-stage recommender systems \cite{yi2019sampling} and RAG technique \cite{lewis2020retrieval}, a LLM is applied on the list of tracks and tags beforehand transformed into an unstructured text, to prioritize tracks that best match the original query. Additional rules such as artists diversity and overall quality of the playlist are also mentioned in the LLM prompt to optimize the final user experience.

\section{Deploying Text2Playlist on Deezer}
\label{deploy}

\subsection{System Deployment}

From a technical standpoint, this system was designed as a stand-alone framework, written in Python and running on a Kubernetes cluster. The LLM used for tags extraction (\ref{tagsandretrieval}) and playlist refinement (\ref{playlistrefinement}) is Gemini Flash 1.5 \cite{team2024gemini}. This choice was mainly driven by cost considerations, as Gemini Flash costs are calculated based on the number of input and output tokens. Therefore, there is no costs incurred when there is no usage, which is beneficial when gradually rolling out such a feature that may not be in constant use. User and song-related data (\ref{tagsandretrieval}), including CF embedding vectors, mood scores and other catalog information, are exported daily in a Cassandra cluster. To retrieve the tracks matching the extracted tags we use Elasticsearch \cite{gormley2015elasticsearch}. For user-song affinity scoring (\ref{leveragingcf}), we use a Golang application incorporating the Faiss library \cite{johnson2019billion}.

\subsection{Empirical Analysis}

% Shortly after the release, several technology-focused articles were published highlighting the launch of the feature. During the playlist creation process, users can choose between manually creating a playlist or utilizing PLAYLIST IA. 

\label{empiricalanalysis}

After weeks of internal tests, Text2Playlist was released in July 2024 to 5\% of premium users on mobile and web. Since October 2024, it has been expanded to 20\% with success, which is promising for its wider roll-out. One key metric we analyze to gauge user satisfaction is the proportion of playlists generated by the feature that are listened to in the following days: while it occurs for 27\% of manual playlists, it is 45\% for the generated playlists by Text2Playlist, indicating a positive engagement with the feature. Lastly, we report most frequent tags of moods asked by the users are “Chill” and “Party”, representing even nearly half of the requested moods. These observations provide valuable insights illustrating what kind of music users want through broad queries on Deezer over time.
%et le nb de track distinct moyen par playlist écouté est de 17 pour classique (contre 22 pour genAI)

Note: our experiments will be further illustrated and discussed in the ECIR "Industry Talk" associated with this article. Resources related to this talk will be available on: \href{https://github.com/deezer/text2playlist-ecir2025}{https://github.com/deezer/text2playlist-ecir2025}.

% Despite users having the freedom to express their preferences in any manner, we observed that the average length of queries was relatively short, often resembling a list of keywords rather than complete sentences. To promote user creativity, we implemented placeholders intended to inspire more descriptive input.

\section{Conclusion}
\label{conclusion}

In this paper, we presented Text2Playlist, a stand-alone tool designed for generating personalized playlists from text at scale. Text2Playlist feature was successfully deployed on the music streaming service Deezer in 2024. Beyond demonstrating promising performance, this system provides valuable insights about users' music needs and how they articulate them (e.g., what are the most popular moods recognized by the LLM in the queries  \ref{empiricalanalysis}?). Our team also plans to increase the coverage and diversity of the extracted tags (e.g., could we use LLM or lyrics to enrich even more music representation \cite{delbouys2018music}?). Besides, as shown in \ref{empiricalanalysis}, users often need to reformulate queries: with the surge of assistant-driven interactions, we could refactor Text2Playlist into a conversational tool \cite{liao2023proactive}.

\section*{Speaker and Company}

\textbf{Mathieu Delcluze} is a Senior Data Scientist at \textbf{Deezer}, a French music streaming service created in 2007 and with over 16~million active users in 180 countries. In the Search team, he develops large-scale machine learning systems to improve music retrieval on this service. Resources related to his talk will be available on: \href{https://github.com/deezer/text2playlist-ecir2025}{https://github.com/deezer/text2playlist-ecir2025}.

\bibliographystyle{splncs04}
\bibliography{ecir25_lb3}

\begin{thebibliography}{10}
\providecommand{\url}[1]{\texttt{#1}}
\providecommand{\urlprefix}{URL }
\providecommand{\doi}[1]{https://doi.org/#1}

\bibitem{bai2024leveraging}
Bai, X., Wu, X., Stojkovic, I., Tsioutsiouliklis, K.: Leveraging large language models for improving keyphrase generation for contextual targeting. In: Proceedings of the 33rd ACM International Conference on Information and Knowledge Management. pp. 4349--4357 (2024)

\bibitem{bainbridge2003people}
Bainbridge, D., Cunningham, S.J., Downie, J.S.: How people describe their music information needs: A grounded theory analysis of music queries  (2003)

\bibitem{bendada2023track}
Bendada, W., Bontempelli, T., Morlon, M., Chapus, B., Cador, T., Bouab{\c{c}}a, T., Salha-Galvan, G.: {Track Mix Generation on Music Streaming Services using Transformers}. In: Proceedings of the 17th ACM Conference on Recommender Systems. pp. 112--115 (2023)

\bibitem{bendada2020carousel}
Bendada, W., Salha, G., Bontempelli, T.: {Carousel Personalization in Music Streaming Apps with Contextual Bandits}. In: Proceedings of the 14th ACM Conference on Recommender Systems. pp. 420--425 (2020)

\bibitem{bokde2015matrix}
Bokde, D., Girase, S., Mukhopadhyay, D.: {Matrix Factorization Model in Collaborative Filtering Algorithms: A Survey}. Procedia Computer Science  \textbf{49},  136--146 (2015)

\bibitem{bontempelli2022flow}
Bontempelli, T., Chapus, B., Rigaud, F., Morlon, M., Lorant, M., Salha-Galvan, G.: {Flow Moods: Recommending Music by Moods on Deezer}. In: Proceedings of the 16th ACM Conference on Recommender Systems. pp. 452--455 (2022)

\bibitem{briand2024let}
Briand, L., Bontempelli, T., Bendada, W., Morlon, M., Rigaud, F., Chapus, B., Bouab{\c{c}}a, T., Salha-Galvan, G.: Let’s get it started: Fostering the discoverability of new releases on deezer. In: European Conference on Information Retrieval. pp. 286--291. Springer (2024)

\bibitem{briand2021semi}
Briand, L., Salha-Galvan, G., Bendada, W., Morlon, M., Tran, V.A.: {A Semi-Personalized System for User Cold Start Recommendation on Music Streaming Apps}. In: Proceedings of the 27th ACM SIGKDD Conference on Knowledge Discovery and Data Mining. pp. 2601--2609 (2021)

\bibitem{choi2020prediction}
Choi, J., Khlif, A., Epure, E.: Prediction of user listening contexts for music playlists. In: Proceedings of the 1st Workshop on NLP for Music and Audio (NLP4MusA). pp. 23--27 (2020)

\bibitem{deezerwebsite}
Deezer: \url{https://www.deezer.com} (2023)

\bibitem{delbouys2018music}
Delbouys, R., Hennequin, R., Piccoli, F., Royo-Letelier, J., Moussallam, M.: Music mood detection based on audio and lyrics with deep neural net. arXiv preprint arXiv:1809.07276  (2018)

\bibitem{epure2020multilingual}
Epure, E.V., Salha, G., Hennequin, R.: Multilingual music genre embeddings for effective cross-lingual music item annotation. arXiv preprint arXiv:2009.07755  (2020)

\bibitem{garcia2018understanding}
Garcia-Gathright, J., St.~Thomas, B., Hosey, C., Nazari, Z., Diaz, F.: Understanding and evaluating user satisfaction with music discovery. In: The 41st International ACM SIGIR Conference on Research \& Development in Information Retrieval. pp. 55--64 (2018)

\bibitem{gong2021ast}
Gong, Y., Chung, Y.A., Glass, J.: Ast: Audio spectrogram transformer. arXiv preprint arXiv:2104.01778  (2021)

\bibitem{gormley2015elasticsearch}
Gormley, C., Tong, Z.: Elasticsearch: the definitive guide: a distributed real-time search and analytics engine. " O'Reilly Media, Inc." (2015)

\bibitem{hennequin2018audio}
Hennequin, R., Royo-Letelier, J., Moussallam, M.: Audio based disambiguation of music genre tags. arXiv preprint arXiv:1809.07256  (2018)

\bibitem{hosey2019just}
Hosey, C., Vujovi{\'c}, L., St.~Thomas, B., Garcia-Gathright, J., Thom, J.: Just give me what i want: How people use and evaluate music search. In: Proceedings of the 2019 chi conference on human factors in computing systems. pp. 1--12 (2019)

\bibitem{ibrahim2020should}
Ibrahim, K.M., Epure, E.V., Peeters, G., Richard, G.: Should we consider the users in contextual music auto-tagging models? In: 21st International Society for Music Information Retrieval Conference (2020)

\bibitem{ibrahim2022exploiting}
Ibrahim, K.M., Epure, E.V., Peeters, G., Richard, G.: Exploiting device and audio data to tag music with user-aware listening contexts. arXiv preprint arXiv:2211.07250  (2022)

\bibitem{ibrahim2020audio}
Ibrahim, K.M., Royo-Letelier, J., Epure, E.V., Peeters, G., Richard, G.: Audio-based auto-tagging with contextual tags for music. In: ICASSP 2020-2020 IEEE International Conference on Acoustics, Speech and Signal Processing (ICASSP). pp. 16--20. IEEE (2020)

\bibitem{issa2023comparative}
Issa, B., Jasser, M.B., Chua, H.N., Hamzah, M.: A comparative study on embedding models for keyword extraction using keybert method. In: 2023 IEEE 13th International Conference on System Engineering and Technology (ICSET). pp. 40--45. IEEE (2023)

\bibitem{johnson2019billion}
Johnson, J., Douze, M., J{\'e}gou, H.: {Billion-Scale Similarity Search with GPUs}. IEEE Transactions on Big Data  \textbf{7}(3),  535--547 (2019)

\bibitem{koren2015advances}
Koren, Y., Bell, R.: {Advances in Collaborative Filtering}. Recommender Systems Handbook pp. 77--118 (2015)

\bibitem{lamkhede2019challenges}
Lamkhede, S., Das, S.: Challenges in search on streaming services: netflix case study. In: Proceedings of the 42nd International ACM SIGIR Conference on Research and Development in Information Retrieval. pp. 1371--1374 (2019)

\bibitem{laplante2008everyday}
Laplante, A.: Everyday life music information-seeking behaviour of young adults: an exploratory study  (2008)

\bibitem{lee2017survey}
Lee, S., Masoud, M., Balaji, J., Belkasim, S., Sunderraman, R., Moon, S.J.: A survey of tag-based information retrieval. International Journal of Multimedia Information Retrieval  \textbf{6},  99--113 (2017)

\bibitem{lewis2020retrieval}
Lewis, P., Perez, E., Piktus, A., Petroni, F., Karpukhin, V., Goyal, N., K{\"u}ttler, H., Lewis, M., Yih, W.t., Rockt{\"a}schel, T., et~al.: Retrieval-augmented generation for knowledge-intensive nlp tasks. Advances in Neural Information Processing Systems  \textbf{33},  9459--9474 (2020)

\bibitem{li2019search}
Li, A., Thom, J., Chandar, P., Hosey, C., Thomas, B.S., Garcia-Gathright, J.: Search mindsets: Understanding focused and non-focused information seeking in music search. In: The World Wide Web Conference. pp. 2971--2977 (2019)

\bibitem{li2017leveraging}
Li, J., Tang, Y., Chen, J.: Leveraging tagging and rating for recommendation: Rmf meets weighted diffusion on tripartite graphs. Physica A: Statistical Mechanics and its Applications  \textbf{483},  398--411 (2017)

\bibitem{liao2023proactive}
Liao, L., Yang, G.H., Shah, C.: Proactive conversational agents in the post-chatgpt world. In: Proceedings of the 46th International ACM SIGIR Conference on Research and Development in Information Retrieval. pp. 3452--3455 (2023)

\bibitem{mao2020item}
Mao, K., Xiao, X., Zhu, J., Lu, B., Tang, R., He, X.: Item tagging for information retrieval: A tripartite graph neural network based approach. In: Proceedings of the 43rd International ACM SIGIR Conference on Research and Development in Information Retrieval. pp. 2327--2336 (2020)

\bibitem{meseguer2024experimental}
Meseguer-Brocal, G., Desblancs, D., Hennequin, R.: An experimental comparison of multi-view self-supervised methods for music tagging. In: ICASSP 2024-2024 IEEE International Conference on Acoustics, Speech and Signal Processing (ICASSP). pp. 1141--1145. IEEE (2024)

\bibitem{naveed2023comprehensive}
Naveed, H., Khan, A.U., Qiu, S., Saqib, M., Anwar, S., Usman, M., Akhtar, N., Barnes, N., Mian, A.: A comprehensive overview of large language models. arXiv preprint arXiv:2307.06435  (2023)

\bibitem{penha2023improving}
Penha, G., Palumbo, E., Aziz, M., Wang, A., Bouchard, H.: Improving content retrievability in search with controllable query generation. In: Proceedings of the ACM Web Conference 2023. pp. 3182--3192 (2023)

\bibitem{petroni2019language}
Petroni, F., Rockt{\"a}schel, T., Lewis, P., Bakhtin, A., Wu, Y., Miller, A.H., Riedel, S.: Language models as knowledge bases? arXiv preprint arXiv:1909.01066  (2019)

\bibitem{sguerra2022navigational}
Sguerra, B., Baranes, M., Hennequin, R., Moussallam, M.: Navigational, informational or punk-rock? an exploration of search intent in the musical domain. In: Proceedings of the 30th ACM Conference on User Modeling, Adaptation and Personalization. pp. 202--211 (2022)

\bibitem{su2018user}
Su, N., He, J., Liu, Y., Zhang, M., Ma, S.: User intent, behaviour, and perceived satisfaction in product search. In: Proceedings of the Eleventh ACM International Conference on Web Search and Data Mining. pp. 547--555 (2018)

\bibitem{team2024gemini}
Team, G., Georgiev, P., Lei, V.I., Burnell, R., Bai, L., Gulati, A., Tanzer, G., Vincent, D., Pan, Z., Wang, S., et~al.: Gemini 1.5: Unlocking multimodal understanding across millions of tokens of context. arXiv preprint arXiv:2403.05530  (2024)

\bibitem{yi2019sampling}
Yi, X., Yang, J., Hong, L., Cheng, D.Z., Heldt, L., Kumthekar, A., Zhao, Z., Wei, L., Chi, E.: Sampling-bias-corrected neural modeling for large corpus item recommendations. In: Proceedings of the 13th ACM conference on recommender systems. pp. 269--277 (2019)

\bibitem{zhao2023survey}
Zhao, W.X., Zhou, K., Li, J., Tang, T., Wang, X., Hou, Y., Min, Y., Zhang, B., Zhang, J., Dong, Z., et~al.: A survey of large language models. arXiv preprint arXiv:2303.18223  (2023)

\bibitem{zhou2023comprehensive}
Zhou, C., Li, Q., Li, C., Yu, J., Liu, Y., Wang, G., Zhang, K., Ji, C., Yan, Q., He, L., et~al.: A comprehensive survey on pretrained foundation models: A history from bert to chatgpt. arXiv preprint arXiv:2302.09419  (2023)

\end{thebibliography}

 \end{document}